\documentclass[12pt]{iopart}

\usepackage{bm}
\usepackage{epsfig}
\usepackage{citesort}
\usepackage{graphicx}
\eqnobysec

\newcommand{\lwig}{\mbox{\;\raisebox{.3ex}
    {$<$}$\!\!\!\!\!$\raisebox{-.9ex}{$\sim$}\;}}

\newcommand{\lambdabar}%
{{\hbox{$\lambda$\kern-1.ex\raise+0.45ex\hbox{--}}}}

\long\def\dump#1{}

\begin{document}


\begin{flushright}
{\large \tt MPP-2007-81}
\end{flushright}

\title{Cosmological constraints on neutrino plus axion
hot dark matter}

\author{S.~Hannestad$^1$, A.~Mirizzi$^2$,
G.~G.~Raffelt$^2$ and Y.~Y.~Y.~Wong$^2$}

\address{$^1$~Department of Physics and Astronomy\\
 University of Aarhus, DK-8000 Aarhus C, Denmark\\
 $^2$~Max-Planck-Institut f\"ur Physik (Werner-Heisenberg-Institut)\\
 F\"ohringer Ring 6, D-80805 M\"unchen, Germany}

\ead{\mailto{sth@phys.au.dk}, \mailto{amirizzi@mppmu.mpg.de},
     \mailto{raffelt@mppmu.mpg.de} and \\
     \mailto{ywong@mppmu.mpg.de}}

\begin{abstract}
We use observations of the cosmological large-scale structure to
derive limits on two-component hot dark matter consisting of
mass-degenerate neutrinos and hadronic axions, both components
having velocity dispersions corresponding to their respective
decoupling temperatures. We restrict the data samples to the safely
linear regime, in particular excluding the Lyman-$\alpha$ forest.
Using standard Bayesian inference techniques we derive credible
regions in the two-parameter space of $m_a$ and $\sum m_\nu$.
Marginalising over $\sum m_\nu$ provides $m_a< 1.2 $~eV (95\% C.L.).
In the absence of axions the same data and methods give $\sum
m_\nu< 0.65 $~eV (95\% C.L.). We also derive limits on $m_a$ for a
range of axion--pion couplings up to one order of magnitude
larger or smaller than the hadronic value.
\end{abstract}

\maketitle

\section{Introduction}                        \label{sec:introduction}

The masses of the lightest known particles (neutrinos) are best
constrained by the largest known scales (the entire universe). The
well-established method of using cosmological precision data to
constrain the cosmic hot dark matter
fraction~\cite{Lesgourgues:2006nd,Hannestad:2006zg} has been
extended to hypothetical low-mass particles, notably to axions, in
several
papers~\cite{Hannestad:2003ye,Hannestad:2005df,Melchiorri:2007cd}.

We return to this topic to extend previous studies by some of
us~\cite{Hannestad:2003ye,Hannestad:2005df} in several ways. First,
we update the cosmological data sets to include the Wilkinson
Microwave Anisotropy Probe 3-year data as well as the baryon
acoustic oscillations measurements from the Sloan Digital Sky Survey
that have since become available. Second, we use standard Bayesian
inference techniques to construct credible regions in parameter
space, in contrast to the likelihood maximisation method used
before~\cite{Hannestad:2003ye,Hannestad:2005df}.  Most importantly,
we consider a two-component hot dark matter fraction consisting of
axions and neutrinos. Since neutrinos are known to have nonvanishing
masses, their hot-dark matter contribution is an unavoidable
cosmological fit parameter. Axions and neutrinos decouple at
different epochs and thus have different velocity dispersions that
we implement self-consistently. In this regard our work parallels a
recent study by another group~\cite{Melchiorri:2007cd}.

We begin in section~\ref{sec:axions} with a brief summary of the
relevant axion parameters and their decoupling conditions. In
section~\ref{sec:model} we describe the cosmological model and the
parameter space we use. In section~\ref{sec:data} we summarise the
included data sets and briefly discuss our reasons for limiting the
analysis to data in the safely linear regime of structure formation.
We derive our new constraints in section~\ref{sec:results} before
concluding in section~\ref{sec:conclusions}.

\section{Hot dark matter axions}                    \label{sec:axions}

The Peccei--Quinn solution of the CP problem of strong interactions
predicts the existence of axions, low-mass pseudoscalars that are
very similar to neutral pions, except that their mass and
interaction strengths are suppressed by a factor of order
$f_\pi/f_a$, where $f_\pi\approx 93$~MeV is the pion decay constant,
and $f_a$ a large energy scale, the axion decay constant or
Peccei--Quinn scale~\cite{Peccei:2006as}. In more detail, the axion
mass~is
\begin{equation}\label{eq:axmass}
 m_a=C_a\,\frac{z^{1/2}}{1+z}\,\frac{f_\pi m_\pi}{f_a}
 =C_a\,\frac{6.0~{\rm eV}}{f_a/10^6~{\rm GeV}}\,,
\end{equation}
where $z=m_u/m_d$ is the mass ratio of the up and down quarks. We
will follow the previous axion literature and assume a value
$z=0.56$ \cite{Gasser:1982ap,Leutwyler:1996qg}, but we note that it
could vary in the range 0.3--0.6 \cite{Yao:2006px}. Because of this
uncertainty and to cover more general cases we will sometimes
include a fudge factor $C_a$ with the standard value~1. We will
consider cases with $-1<\log_{10}(C_a)<+1$.

A large range of $f_a$ values (or, equivalently, $m_a$ values) can be
excluded by experiments and by astrophysical and cosmological
arguments~\cite{Raffelt:2006rj}. Axions with a mass of order
10~$\mu$eV could well be the cold dark matter of the
universe~\cite{Sikivie:2006ni} and if so will be found
eventually by the ongoing ADMX experiment provided that $1~\mu{\rm
eV}<m_a<100~\mu{\rm eV}$~\cite{Asztalos:2006kz}. In addition, a hot
axion population is produced by thermal
processes~\cite{Chang:1993gm,Turner:1986tb,Masso:2002np}. Axions
attain thermal equilibrium at the QCD phase transition or later if
$f_a\lwig10^8$~GeV, erasing the cold axion population produced
earlier and providing a hot dark matter component instead.

In principle, $f_a\lwig 10^9$~GeV is excluded by the supernova SN~1987A
neutrino burst duration~\cite{Raffelt:2006rj}. However, the sparse data
sample, our poor understanding of the nuclear medium in the supernova
interior, and simple prudence suggest that one should not base
far-reaching conclusions about the existence of axions in this
parameter range on a single argument or experiment alone. Therefore,
it remains important to tap other sources of information, especially
if they are easily available.

For those axion models with nonvanishing couplings to charged
fermions, there exist stellar energy loss limits based on the axion--electron
coupling that are competitive with the SN~1987A constraints so
that here one does not rely on a single argument to exclude axions
in the $f_a\lwig 10^9$~GeV range. Therefore, we focus on hadronic
models where axions do not directly couple to ordinary quarks and
leptons. In this class of models all axion properties depend on
$f_a$ alone and not on model-dependent Peccei--Quinn charges of the
ordinary quarks and leptons.

If axions do not couple to charged leptons, the main thermalisation
process in the post-QCD epoch is~\cite{Chang:1993gm}
\begin{equation}\label{eq:pionprocess}
a+\pi\leftrightarrow\pi+\pi\,.
\end{equation}
The axion--pion interaction is given by a Lagrangian of the
form~\cite{Chang:1993gm}
\begin{equation}\label{eq:axionpionlagrangian}
{\cal L}_{a\pi}=\frac{C_{a\pi}}{f_\pi f_a}\,
\left(\pi^0\pi^+\partial_\mu\pi^- +\pi^0\pi^-\partial_\mu\pi^+
-2\pi^+\pi^-\partial_\mu\pi^0\right)
\partial_\mu a\,.
\end{equation}
In hadronic axion models, the coupling constant
is~\cite{Chang:1993gm}
\begin{equation}\label{eq:axionpioncoupling}
C_{a\pi}=\frac{1-z}{3\,(1+z)}\,.
\end{equation}
We note that in general the chiral symmetry-breaking Lagrangian
gives rise to an additional piece for ${\cal L}_{a\pi}$
proportional to $(m_\pi^2/f_\pi
f_a)\,(\pi^0\pi^0+2\pi^-\pi^+)\ \pi^0a$. However, for hadronic axion
models this term vanishes identically, in contrast, for example, to
the DFSZ model (Roberto Peccei, private communication).

Based on the axion--pion interaction, the axion decoupling
temperature in the early universe was calculated by some of us in
Ref.~\cite{Hannestad:2005df}, where all relevant details are
reported. In our standard case we use the axion mass $m_a$ as our
primary parameter from which we derive the corresponding axion--pion
interaction strength by virtue of equations~(\ref{eq:axmass})
and~(\ref{eq:pionprocess}). Noting that even in hadronic axion
models there is some uncertainty in this relationship due to the
uncertain quark-mass ratio $z$, we consider also more general cases
in which we include a fudge factor $C_a$ as defined in equation~(\ref{eq:axmass}),
thus allowing for a more general relationship between $m_a$ and
$C_{a\pi}$.

\section{Cosmological model}                         \label{sec:model}

We consider a cosmological model with vanishing spatial curvature and
adiabatic initial conditions, described by nine free parameters,
\begin{equation}\label{eq:model}
{\bm \theta} = \{\omega_{\rm dm},\omega_b,H_0,\tau,
\ln(10^{10}A_s),n_s,
\sum m_\nu,m_a,\log_{10}(C_a)\}.
\end{equation}
Here, $\omega_{\rm dm}=\Omega_{\rm
dm} h^2$ is the physical dark matter density, $\omega_b=\Omega_b h^2$ the baryon density,
$H_0=h~100~{\rm km~s^{-1}~Mpc^{-1}}$ the Hubble
parameter, $\tau$ the optical depth to reionisation,
$A_s$ the amplitude of the primordial scalar power spectrum,
and $n_s$ its spectral index.  These six parameters represent the
simplest parameter set necessary for a consistent interpretation
of the currently available data.

In addition, we allow for a nonzero sum of neutrino masses $\sum
m_\nu$, a nonvanishing axion mass $m_a$, and a fudge factor $C_a$
relating $m_a$ to $f_a$ as defined in equation~(\ref{eq:axmass}). These
extra parameters will be varied one at a time, as well as in
combination. Their ``standard'' values are given in
table~\ref{tab:priors}, along with the priors for all cosmological
fit parameters considered here.

\begin{table}
\caption{Priors and standard values for the cosmological fit
parameters considered in this work.  All priors are
uniform in the given intervals (i.e., top hat).\label{tab:priors}} \hskip25mm
{\footnotesize
\begin{tabular}{lll}
\br
Parameter& Standard & Prior\\
\mr
$\omega_{\rm dm}$   & ---   & $0.01$--$0.99$ \\
$\omega_{\rm b}$    & ---   & $0.005$--$0.1$ \\
$h$                 & ---   & $0.4$--$1.0$\\
$\tau$              & ---   & $0.01$--$0.8$ \\
$\ln(10^{10}A_s)$   & ---   & $2.7$--$4.0$ \\
$n_s$               & ---   & $0.5$--$1.5$ \\
$\sum m_\nu$ [eV]       & 0 & $0$--$20$ \\
$m_a$ [eV]      & 0     & $0$--$20$ \\
$\log_{10}(C_a)$    & 0 & $-1$--$1$ \\
\br
\end{tabular}
}
\end{table}

\section{Data}                                        \label{sec:data}

\subsection{Cosmic microwave background (CMB)}

We use CMB data from the Wilkinson Microwave Anisotropy Probe (WMAP)
ex\-per\-i\-ment after three years of observation
\cite{Spergel:2006hy,Hinshaw:2006ia,Page:2006hz}. The data analysis
is performed using version~2 of the likelihood calculation package
provided by the WMAP team on the LAMBDA homepage \cite{lambda}.

\subsection{Large scale structure (LSS)}

We use the large-scale galaxy power spectra $P_{\rm g}(k)$
inferred from the luminous red galaxy (LRG) sample
of the Sloan Digital Sky Survey (SDSS)
\cite{Percival:2006gt,Tegmark:2006az}
and from the Two-degree Field Galaxy Redshift Survey
(2dF)~\cite{Cole:2005sx}.  These power spectra are related to
the underlying matter power spectrum $P_{\rm m}(k)$ via
 $P_{\rm g}(k)=b^2(k) P_{\rm m}(k)$, where the galaxy bias $b(k)$
is conventionally assumed to be constant with respect to $k$ over
the scales probed by galaxy clustering surveys.

However, recent studies suggest that this assumption may break down
beyond $k\sim 0.1 \ h \ {\rm Mpc}^{-1}$, and may source
the apparent tension between the SDSS and the 2dF-inferred galaxy power spectra~%
\cite{Percival:2006gt,Cole:2006kn}.  To model the effects of
scale-dependent biasing when extracting cosmological
parameters  from galaxy clustering data, both the SDSS-LRG and the 2dF teams advocate
the use of the $Q_{\rm nl}$ fitting formula developed in
Ref.~\cite{Cole:2005sx} for $\Lambda$CDM cosmologies.
See Ref.~\cite{Hamann:2007pi} for a detailed discussion.

In the present work, however, we take the view that
fitting formulae developed for standard cosmologies may not be
applicable in nonstandard scenarios, particularly those involving
new length scales arising from, e.g., axion and neutrino
free-streaming.
Developing an alternative formula to properly handle these nonstandard
effects on the galaxy bias
is also beyond our present scope.
We therefore adopt a conservative approach, and use only power spectrum data
well below  $k\sim 0.1 \ h \ {\rm Mpc}^{-1}$, where a  scale-{\it independent}
bias is likely to hold true:
\begin{itemize}
\item 2dF, $k_{\rm max} \sim 0.09 \ h \ {\rm Mpc}^{-1}$ (17 bands),
\item SDSS-LRG, $k_{\rm max} \sim 0.07 \ h \ {\rm Mpc}^{-1} $ (11 bands).
\end{itemize}
The combined set of these data is denoted LSS. We assume a scale-independent
bias for each data set, and marginalise analytically over each
bias parameter $b^2$ with a flat prior.

\subsection{Baryon acoustic oscillations (BAO)}

The baryon acoustic oscillations peak has been measured in the SDSS
luminous red galaxy sample \cite{Eisenstein2005}.  We use all
20~points in the two-point correlation data set supplied in
Ref.~\cite{Eisenstein2005} and the analysis procedure described
therein, including power spectrum dewiggling, nonlinear corrections
with the {\sc Halofit} package~\cite{halofit}, corrections for
redshift-space distortion, and analytic marginalisation over the
normalisation of the correlation function.

\subsection{Type Ia supernovae (SNIa)}

We use the luminosity distance measurements of distant type~Ia
supernovae provided by Davis et~al.~\cite{Davis:2007na}. This sample
is a compilation of supernovae measured by the Supernova Legacy
Survey (SNLS) \cite{Astier:2005qq}, the ESSENCE project
\cite{Wood-Vasey:2007jb}, and the Hubble Space Telescope
\cite{Riess:2006fw}, as well as a set of 45 nearby supernovae. In
total the sample contains 192 supernovae.

\subsection{Lyman-$\alpha$ forest (Ly$\alpha$)}
\label{sec:lya}

Measurements of the flux power spectrum of the Lyman-$\alpha$ forest
has been used to reconstruct the matter power spectrum on small scales
at large redshifts. By far the largest sample of spectra comes from
the SDSS survey. This data set was carefully analysed in
McDonald et~al.~\cite{mcdonald} and used to constrain the linear matter
power spectrum. The derived linear fluctuation amplitude at
$k=0.009~{\rm km~s}^{-1}$ and $z=3$ is
$\Delta^2 = 0.452^{+0.07}_{-0.06}$, and the effective spectral index
 $n_{\rm eff} = -2.321^{+0.06}_{-0.05}$. These results were
derived using a very elaborate model of the local intergalactic
medium in conjunction with hydrodynamic simulations.

While the Ly$\alpha$ data provide in principle a very powerful
probe of the fluctuation amplitude on small scales, the question
remains as to the level of systematic uncertainty in the result. The
same data have been reanalysed by Seljak et~al.\ \cite{Seljak:2006bg}
and Viel et~al.\ \cite{Viel:2005eg,Viel:2005ha,viel2006}, with
somewhat different results. Specifically, the normalisation found in
Refs.~\cite{Viel:2005eg,Viel:2005ha,viel2006} is lower than that
reported in Ref.~\cite{mcdonald}.

This question of normalisation is particularly important for bounds
on the hot dark matter content of the universe. Since the
free-streaming scale of light neutrinos or axions is larger than the
length scale probed by Ly$\alpha$, their effect on the
Ly$\alpha$ data amounts to an overall change in the
normalisation that is completely degenerate with any possible shift
due to systematics. The Ly$\alpha$ analysis in Ref.~\cite{mcdonald}
already points to a higher fluctuation amplitude $\Delta^2$ than that
derived from the WMAP 3-year data; the addition of a hot dark matter
component will render the two data sets even less compatible.
This incompatibility in turn leads to a much stronger formal
bound on the mass of the hot dark matter particle than would
be expected considering the sensitivity of the present data
(this is true for both neutrinos and other types of hot dark matter,
such as axions).

These considerations suggest that the Ly$\alpha$ data are
at present dominated by systematic effects.   We therefore
refrain from using them in the present analysis.

\begin{table}
\caption{1D marginal 95\% upper bounds on  $\sum m_\nu$ and $m_a$ for
several different choices of data sets and models.
\label{tab:fudge}} \hskip25mm {\footnotesize
\begin{tabular}{lccc}
\br
Data set & $C_a$ prior & $\sum m_\nu$ [eV] & $m_a$ [eV]\\
\mr
WMAP+LSS+SNIa & $\log_{10}(C_a)=0$&  0.63  & 2.0  \\
WMAP+LSS+SNIa+BAO & & 0.59  & 1.2  \\
\ms
Fixed $\sum m_\nu=0$ \\
WMAP+LSS+SNIa+BAO & & --- & 1.4 \\
\ms
Fixed $m_a=0$ \\
WMAP+LSS+SNIa+BAO & & 0.65 & --- \\
\mr
WMAP+LSS+SNIa &  $-1 < \log_{10}(C_a) < 1$ & 0.61  & 2.2  \\
WMAP+LSS+SNIa+BAO &  & 0.60  & 1.1  \\
\br
\end{tabular}
}
\end{table}

\section{Results}                                  \label{sec:results}

\begin{figure}
\hspace{25mm}
\includegraphics[width=10.cm]{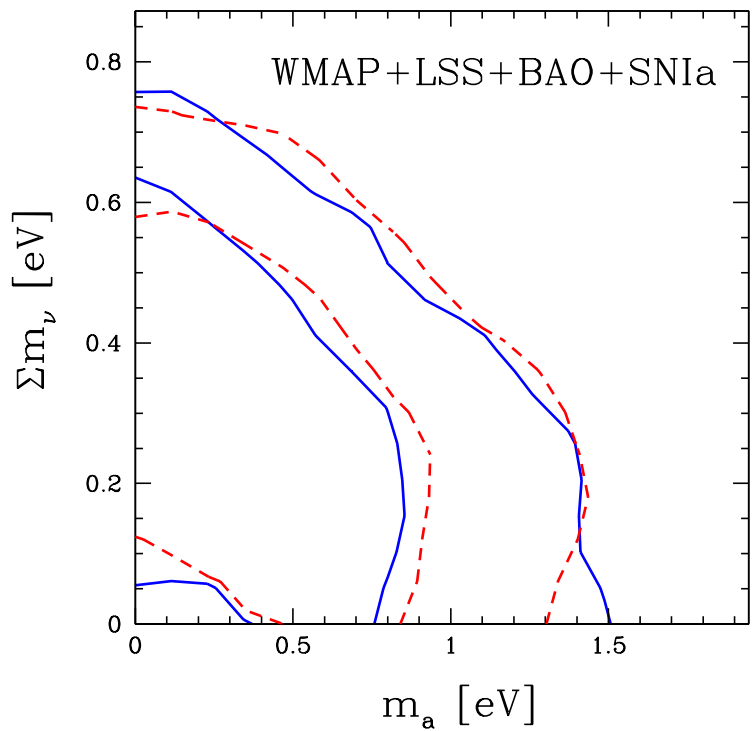}
\caption{2D marginal 68\% and 95\% contours in the $\sum m_\nu$-$m_a$ plane
derived
from the full data set WMAP+LSS+BAO+SNIa.
The blue/solid lines correspond to the fudge factor being fixed at
$\log_{10}(C_a)=0$, while the
red/dashed lines indicate a top-hat prior on $C_a$ in the interval  $-1<\log_{10}(C_a)<1$.
\label{fig:mamnu}}
\end{figure}

\begin{figure}
\hspace{25mm}
\includegraphics[width=10.cm]{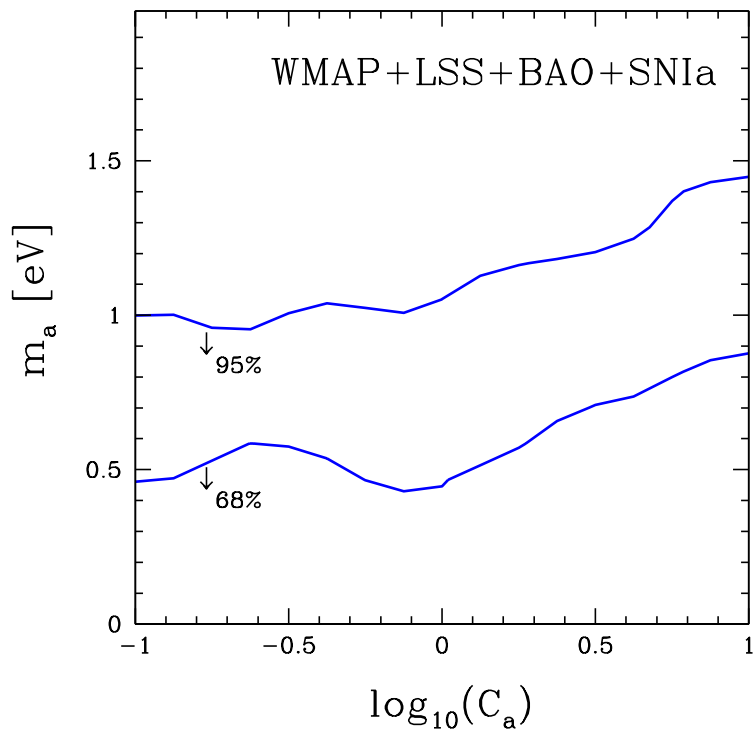}
\caption{2D marginal 68\% and 95\% contours in the $m_a$-$\log_{10}(C_a)$ plane,
assuming a top-hat prior on the fudge factor in the interval $-1<\log_{10}(C_a)<1$.
\label{fig:mafudge}}
\end{figure}

We use standard Bayesian inference techniques, and explore the model parameter space
with Monte Carlo Markov Chains (MCMC) generated using
the publicly available {\sc CosmoMC}
package~\cite{Lewis:2002ah,cosmomc}.  Our results are summarised in
tables~\ref{tab:fudge} and \ref{tab:fudge1D}, and
figures~\ref{fig:mamnu} and \ref{fig:mafudge}.

For the case of a standard hadronic axion our bounds on $\sum m_\nu$
and $m_a$ are tabulated in table~\ref{tab:fudge}. When BAO data are
included our bounds are almost identical to those
recently derived in Ref.~\cite{Melchiorri:2007cd} based on their conservative
data set.
The main difference is that we do not use the HST prior on
$h$, but instead include the SDSS-BAO data. Since the BAO data break
the $\Omega_m$-$h$ degeneracy,  their inclusion has much the same
effect as adding the HST prior. The importance of BAO data for the
bound can be seen by the fact that the 95\% upper bound is reduced
from $2.0$~eV to $1.2$~eV. Our complete SNIa data set is also
somewhat larger than the SNLS data set used in
Ref.~\cite{Melchiorri:2007cd}, containing in addition data from the
GOODS and ESSENCE surveys. However, this has only a very modest
impact on our results.

Our neutrino mass bound $\sum m_\nu<0.65$~eV (95\% C.L.) in the absence of axions
is identical to that derived by some of us in
Ref.~\cite{Hannestad:2005df}, whereas the axion mass limit $m_a<1.4$~eV
in the absence of neutrino masses found here is significantly weaker than
the $1.05$~eV limit found earlier~\cite{Hannestad:2005df}. The agreement of the
neutrino mass limits is coincidental because here we use different
data, notably excluding the Lyman-$\alpha$ forest,
and a different statistical methodology
(marginalisation instead of maximisation). The relative
difference between the limits can be interpreted such that the axion
bound benefits more from the inclusion of small-scale data than the
neutrino mass bound, presumably because axions freeze out earlier
and thus have a smaller velocity dispersion. Including the
Ly$\alpha$ data here would strongly improve both limits as can be
gleaned from the results of Ref.~\cite{Melchiorri:2007cd}. Adding
Ly$\alpha$ to their conservative data set, the analysis of
Ref.~\cite{Melchiorri:2007cd} finds that
the marginalised
axion mass limit improves by a factor 0.30, whereas the marginalised
neutrino mass limit improves only by a factor 0.36, i.e., the
relative gain for axions is 20\% stronger. The changes in our new
limits relative to those of Ref.~\cite{Hannestad:2005df} are in
agreement with this picture.

Returning to our new limits, an important observation is that the
upper bound on the sum of neutrino masses is largely independent of
whether or not massive axions are present. The 95\% upper limit on
$\sum m_\nu$ is in either case approximately $0.6$~eV, a bound very
close to that found in previous studies using roughly the same data
combination
\cite{Zunckel:2006mt,Fogli:2006yq,Cirelli:2006kt,Goobar:2006xz}.

In figure 2 we show how the bound on the axion mass changes as $C_a$
is allowed to vary up or down by up to a factor of~10, assuming a
uniform prior on $\log_{10}(C_a)$ between $-1$ and $+1$. Note that
the figure shows the 2D marginal contours, i.e., $C_a$ and $m_a$ are
fitted simultaneously. For $\log_{10}(C_a) \leq 0$, the bound on
$m_a$ does not depend on $C_a$ because the number of degrees of
freedom at decoupling, $g_*(T_D)$, is approximately constant for a
large range of $f_a$ values (see table 2 of
Ref.~\cite{Hannestad:2005df}). For $\log_{10}(C_a)
> 0$,  the value of $g_*(T_D)$ increases significantly with increasing
$C_a$ for a given $m_a$.  This increase in $g_*(T_D)$ leads to a
corresponding drop in the present axion number density,
\begin{equation}
n_a = \frac{g_*({\rm today})}{g_*(T_D)} \times \frac{n_\gamma}{2}.
\end{equation}
 For a fixed $m_a$ this amounts to a
decrease in the ratio $\Omega_a/\Omega_m$ with increasing $C_a$.
The bound on $m_a$ therefore becomes correspondingly weaker.

As can be seen in figure~\ref{fig:mafudge}, the 2D  marginal 95\%  upper limit
on $m_a$ stays roughly constant at \mbox{$m_a \lwig 1.1$~eV} when $\log_{10}(C_a) \leq 0$,
and increases roughly linearly with $\log_{10}(C_a)$ to about $1.4$~eV at
$\log_{10}(C_a) = 1$.  We stress again that the $m_a$ bounds in this figure
are 2D bounds, and are formally---and often also in practice---not equivalent
to 1D bounds on $m_a$ derived under the assumption of a fixed $C_a$.
For instance, the 2D  bound on $m_a$ at $C_a=1$ in figure~\ref{fig:mafudge}
is not exactly identical to the 1D bound quoted in table~\ref{tab:fudge}
for a fixed $C_a=1$.
However, despite this formality, the  $m_a$-$\log_{10}(C_a)$ trend observed
in figure~\ref{fig:mafudge} is also evident
 in table~\ref{tab:fudge1D}, which
shows the 1D marginal 68\% and 95\% bounds on $m_a$ for {\it fixed} values of $C_a$.

Finally, we note that if $C_a$ is increased much beyond 10, the bound
will deteriorate rapidly because axion decoupling will have occurred beyond
the QCD phase transition.

\begin{table}
\caption{1D marginal 68\%/95\% upper bounds on $\sum m_\nu$ and $m_a$
for  fixed fudge
factors $C_a$. The data set used is WMAP+LSS+SNIa+BAO.
\label{tab:fudge1D}} \hskip25mm {\footnotesize
\begin{tabular}{lcc}
\br
$\log_{10}(C_a)$ & $\sum m_\nu$ [eV] & $m_a$ [eV]\\
\mr
$-1.0$  & 0.39/0.64     & 0.51/0.98   \\
$0.0$   & 0.37/0.59     & 0.60/1.2   \\
$1.0$   & 0.40/0.63 & 0.69/1.4 \\
\br
\end{tabular}
}
\end{table}

\section{Conclusions}                           \label{sec:conclusions}

We have updated previous limits from cosmological structure
formation on the mass of hot dark matter axions. This limit applies
to axions which were thermalised, mainly by axion--pion
interactions, in the early universe, and which subsequently
decoupled from the thermal plasma while still relativistic.

In the present study we investigate both the case where the
neutrinos can be regarded as massless, as well as the case in which
massive neutrinos are also allowed to contribute significantly to
the hot dark matter fraction. In both cases we find an upper 95\%
limit on the mass of hadronic axions of $1.1$--$1.2$~eV when all
available cosmological data, except the Lyman-$\alpha$ forest, are
used. Reassuringly, we find that the bound on the sum of neutrino
masses is almost completely unaffected by the presence of hot dark
matter axions.

Because of the uncertainty in the relation between the axion mass
$m_a$ and the energy scale $f_a$, we have also studied the case in
which the relation $m_a = 6.0 \,\ {\rm eV}/(f_a/10^6 \, {\rm GeV})$ is
modified by a fudge factor $C_a$. We have studied $C_a$ in the range
$0.1$--$10$, which is fairly representative of the model
uncertainties. We find that the axion mass bound is largely stable
with respect to varying $C_a$ in this range. For fixed values of
$C_a$, the 1D marginal bound on $m_a$ goes from 0.98~eV at $C_a=0.1$
to 1.4~eV at $C_a=10$. Essentially, this means that for hadronic
axion models the uncertainty of the light quark mass ratios have a
negligible impact on the axion mass limit.

Experimental and astrophysical limits on $m_a$ or $f_a$ are always
derived from limits on the axion coupling to different particles.
The cosmological hot dark matter limit, in contrast, primarily
constrains the axion mass, with a very weak dependence on the
axion--pion coupling. The hot dark matter limit of $m_a\lwig1$~eV is
very similar to the limit derived from globular cluster stars based
on the axion--photon coupling. However, this coupling is quite
uncertain even in hadronic models because even there it depends on
the unknown electric charge of the heavy quark in KSVZ-type models.
The hot dark matter limit implies that it is very difficult to
escape the limit $m_a\lwig1$~eV. One consequence is that in typical
models, axions in the remaining allowed mass range necessarily
escape freely from a supernova core.  By courtesy of the SN~1987A
neutrino burst duration, it follows that one can advance by another
rung in the ladder of different limits and conclude that
$m_a\lwig10^{-2}$~eV \cite{Raffelt:2006rj}. While this SN~1987A
energy-loss limit does not have an obvious loophole, we repeat that
it is based on a very small sample of detected neutrinos and is
subject to various nuclear-physics and axion-model uncertainties.

Our results largely agree with those of
Ref.~\cite{Melchiorri:2007cd} for their conservative data set. In
contrast to Ref.~\cite{Melchiorri:2007cd} and to a previous study by
some of us~\cite{Hannestad:2005df}, we have not included the
Lyman-$\alpha$ forest data which could formally improve both
the neutrino and axion mass limits roughly by a factor of~3. We have explained in
section~\ref{sec:lya} that using the Lyman-$\alpha$ forest exposes one to the
risk that large systematic uncertainties in the normalisation of the
power spectrum at small scales may dominate the final result.

The CAST experiment at CERN searches for axion-like particles
emitted by the Sun by virtue of their coupling to
photons~\cite{Zioutas:2004hi,Andriamonje:2007ew}. By including a helium
filling of the magnet bores with variable pressure one can ``adjust
the photon mass,'' thereby allowing one to probe realistic combinations of
$m_a$ and axion--photon coupling. The completed runs with $^4$He
filling have already extended the experimental sensitivity to $m_a
\sim 0.4$~eV.  Further extensions to up to $m_a \sim 1.16$~eV with
the forthcoming $^3$He runs over three years are on the
agenda~\cite{CAST2007}. This search range is not excluded by our
limits, particularly as we believe that more restrictive limits
derived from the Lyman-$\alpha$ forest may be dominated by
systematic effects that are not reliably controlled.

One further caveat is that limits inferred from cosmological
observations are by and large model-dependent. Additional free
parameters not considered in this work, such as a nonstandard dark
energy equation of state parameter, running in the primordial scalar
spectral index, or a nonzero component of isocurvature modes in the
initial conditions, could conceivably loosen the axion mass bound, as
they have done many times before for the neutrino mass limit
\cite{Zunckel:2006mt,Goobar:2006xz,Hannestad:2005gj}. A significant
and reliable improvement of cosmological hot dark matter limits is
not immediately forthcoming. However, once data from the Planck CMB
experiment \cite{planck} combined with other probes such as weak
lensing surveys of galaxies
\cite{Cooray:1999rv,Song:2003gg,Hannestad:2006as} or of 21-cm
emissions \cite{Metcalf:2006ji}, or high-redshift galaxy surveys
\cite{Takada:2005si,Hannestad:2007cp} become available, the
sensitivity will be pushed down by as mush as an order of magnitude
even in the face of more complicated cosmological model frameworks.
In that event, a detection of axions by CAST in the vicinity of $m_a
\sim 1$~eV will have important ramifications for observational
cosmology.

\section*{Acknowledgements}

We acknowledge use of computing resources from the Danish Center for
Scientific Computing (DCSC) and partial support by the European
Union under the ILIAS project (contract No.\ RII3-CT-2004-506222),
by the Deutsche Forschungsgemeinschaft under the grant TR-27
``Neutrinos and Beyond'' and by The Cluster of Excellence for
Fundamental Physics ``Origin and Structure of the Universe''
(Garching and Munich). A.M.\ is supported by a grant of the
Alexander von Humboldt Foundation. S.H.\ acknowledges support from
the Alexander von Humboldt Foundation through a Friedrich Wilhelm
Bessel Award.

\section*{References}


\begin{thebibliography}{99}

\bibitem{Lesgourgues:2006nd}
  J.~Lesgourgues and S.~Pastor,
  ``Massive neutrinos and cosmology,''
  Phys.\ Rept.\  {\bf 429} (2006) 307
  [arXiv:astro-ph/0603494].

\bibitem{Hannestad:2006zg}
  S.~Hannestad,
  ``Primordial neutrinos,''
  Ann.\ Rev.\ Nucl.\ Part.\ Sci.\  {\bf 56} (2006) 137
  [arXiv:hep-ph/0602058].

\bibitem{Hannestad:2003ye}
  S.~Hannestad and G.~Raffelt,
  ``Cosmological mass limits on neutrinos, axions, and other light
  particles,''
  JCAP {\bf 0404} (2004) 008
  [arXiv:hep-ph/0312154].

\bibitem{Hannestad:2005df}
  S.~Hannestad, A.~Mirizzi and G.~Raffelt,
  ``New cosmological mass limit on thermal relic axions,''
  JCAP {\bf 0507} (2005) 002
  [arXiv:hep-ph/0504059].

\bibitem{Melchiorri:2007cd}
  A.~Melchiorri, O.~Mena and A.~Slosar,
  ``An improved cosmological bound on the thermal axion mass,''
  arXiv:0705.2695 [arXiv:astro-ph].


\bibitem{Peccei:2006as}
  R.~D.~Peccei,
  ``The strong CP problem and axions,''
  submitted to Lecture Notes in Physics
  [arXiv:hep-ph/0607268].

\bibitem{Gasser:1982ap}
  J.~Gasser and H.~Leutwyler,
  ``Quark masses,''
  Phys.\ Rept.\ {\bf 87} (1982) 77.

\bibitem{Leutwyler:1996qg}
  H.~Leutwyler,
  ``The ratios of the light quark masses,''
  Phys.\ Lett.\ B {\bf 378} (1996) 313
  [arXiv:hep-ph/9602366].

\bibitem{Yao:2006px}
  W.~M.~Yao {\it et al.}  [Particle Data Group],
  ``Review of particle physics,''
  J.\ Phys.\ G {\bf 33} (2006) 1.

\bibitem{Raffelt:2006rj}
  G.~G.~Raffelt,
  ``Axions: Motivation, limits and searches,''
  J.\ Phys.\ A: Math.\ Theor.\ 40 (2007) 6607
  [arXiv:hep-ph/0611118].

\bibitem{Sikivie:2006ni}
  P.~Sikivie,
  ``Axion cosmology,''
  submitted to Lecture Notes in Physics
  [arXiv:astro-ph/0610440].

\bibitem{Asztalos:2006kz}
  S.~J.~Asztalos, L.~J.~Rosenberg, K.~van Bibber, P.~Sikivie and K.~Zioutas,
  ``Searches for astrophysical and cosmological axions,''
  Ann.\ Rev.\ Nucl.\ Part.\ Sci.\  {\bf 56} (2006) 293.

\bibitem{Chang:1993gm}
  S.~Chang and K.~Choi,
  ``Hadronic axion window and the big bang nucleosynthesis,''
  Phys.\ Lett.\ B {\bf 316} (1993) 51
  [arXiv:hep-ph/9306216].

\bibitem{Turner:1986tb}
  M.~S.~Turner,
  ``Thermal production of not so invisible axions in
  the early universe,''
  Phys.\ Rev.\ Lett.\  {\bf 59} (1987) 2489
  [Erratum-ibid.\  {\bf 60} (1988) 1101].

\bibitem{Masso:2002np}
  E.~Mass\'o, F.~Rota and G.~Zsembinszki,
  ``On axion thermalization in the early universe,''
  Phys.\ Rev.\ D {\bf 66} (2002) 023004
  [arXiv:hep-ph/0203221].


\bibitem{Spergel:2006hy}
  D.~N.~Spergel {\it et al.}, ``Wilkinson Microwave Anisotropy Probe
  (WMAP) three year results: Implications for cosmology,''
Astrophys.\ J.\ Suppl.\ {\bf 170} (2007) 377
  [arXiv:astro-ph/0603449].

\bibitem{Hinshaw:2006ia}
  G.~Hinshaw {\it et al.},
  ``Three-year Wilkinson Microwave Anisotropy Probe (WMAP)
  observations: Temperature analysis,''
Astrophys.\ J.\ Suppl.\ {\bf 170} (2007) 288
  [arXiv:astro-ph/0603451].

\bibitem{Page:2006hz}
  L.~Page {\it et al.},
  ``Three year Wilkinson Microwave Anisotropy Probe (WMAP)
  observations: Polarization analysis,''
Astrophys.\ J.\ Suppl.\ {\bf 170} (2007) 335
  [arXiv:astro-ph/0603450].

\bibitem{lambda}
  Legacy Archive for Microwave Background Data Analysis
  (LAMBDA), \newline {\tt http://lambda.gsfc.nasa.gov}

\bibitem{Percival:2006gt}
  W.~J.~Percival {\it et al.},
  ``The shape of the SDSS DR5 galaxy power spectrum,''
Astrophys.\ J.\ {\bf 657} (2007) 645
  [arXiv:astro-ph/0608636].

\bibitem{Tegmark:2006az}
  M.~Tegmark {\it et al.},
  ``Cosmological constraints from the SDSS luminous red galaxies,''
  Phys.\ Rev.\  D {\bf 74} (2006) 123507
  [arXiv:astro-ph/0608632].

\bibitem{Cole:2005sx}
  S.~Cole {\it et al.} [2dFGRS Collaboration],
  ``The 2dF Galaxy Redshift Survey: Power-spectrum analysis
  of the final dataset and cosmological implications,''
  Mon.\ Not.\ Roy.\ Astron.\ Soc.\  {\bf 362} (2005) 505
  [arXiv:astro-ph/0501174].

\bibitem{Cole:2006kn}
  S.~Cole, A.~G.~Sanchez and S.~Wilkins,
  ``The galaxy power spectrum: 2dFGRS-SDSS tension?,''
  arXiv:astro-ph/0611178.

\bibitem{Hamann:2007pi}
  J.~Hamann, S.~Hannestad, G.~G.~Raffelt and Y.~Y.~Y.~Wong,
  ``Observational bounds on the cosmic radiation density,''
  arXiv:0705.0440 [arXiv:astro-ph].

\bibitem{Eisenstein2005}
  D.~J.~Eisenstein {\it et al.} [SDSS Collaboration],
  ``Detection of the baryon acoustic peak in the
  large-scale correlation function of SDSS
  luminous red galaxies,''
  Astrophys.\ J.\  {\bf 633} (2005) 560
  [arXiv:astro-ph/0501171];
  see also
  {\tt http://cmb.as.arizona.edu/$\sim$eisenste/acousticpeak}

\bibitem{halofit}
 R.~E.~Smith {\it et al.} [Virgo Consortium Collaboration],
  ``Stable clustering, the halo model and nonlinear cosmological
  power spectra,''
  Mon.\ Not.\ Roy.\ Astron.\ Soc.\ {\bf 341} (2003) 1311
  [arXiv:astro-ph/0207664].

\bibitem{Davis:2007na}
  T.~M.~Davis {\it et al.},
  ``Scrutinizing exotic cosmological models using ESSENCE
  supernova data combined with other cosmological probes,''
  arXiv:astro-ph/0701510.

\bibitem{Astier:2005qq}
  P.~Astier {\it et al.},
  ``The Supernova Legacy Survey: Measurement of $\Omega_M$,
  $\Omega_\Lambda$ and $w$ from the first year data set,''
  Astron.\ Astrophys.\  {\bf 447} (2006) 31
  [arXiv:astro-ph/0510447].

\bibitem{Wood-Vasey:2007jb}
  W.~M.~Wood-Vasey {\it et al.},
  ``Observational constraints on the nature of the dark energy:
  First cosmological results from the ESSENCE supernova survey,''
  arXiv:astro-ph/0701041.

\bibitem{Riess:2006fw}
  A.~G.~Riess {\it et al.},
  ``New Hubble Space Telescope discoveries of type Ia supernovae at $z > 1$:
  Narrowing constraints on the early behavior of dark energy,''
Astrophys.\ J.\ {\bf 659} (2007) 98
  [arXiv:astro-ph/0611572].

\bibitem{mcdonald}
  P.~McDonald {\it et al.},
  ``The linear theory power spectrum from the Lyman-alpha
  forest in the Sloan Digital Sky Survey,''
  Astrophys.\ J.\  {\bf 635} (2005) 761
  [arXiv:astro-ph/0407377].

\bibitem{Seljak:2006bg}
  U.~Seljak, A.~Slosar and P.~McDonald,
  ``Cosmological parameters from combining the Lyman-alpha forest
  with CMB, galaxy clustering and SN constraints,''
  JCAP {\bf 0610} (2006) 014
  [arXiv:astro-ph/0604335].

\bibitem{Viel:2005eg}
  M.~Viel, M.~G.~Haehnelt and V.~Springel,
  ``Testing the accuracy of the Hydro-PM approximation in
  numerical simulations of the Lyman-alpha forest,''
  Mon.\ Not.\ Roy.\ Astron.\ Soc.\  {\bf 367} (2006) 1655
  [arXiv:astro-ph/0504641].

\bibitem{Viel:2005ha}
  M.~Viel and M.~G.~Haehnelt,
  ``Cosmological and astrophysical parameters from the SDSS flux power spectrum
  and hydrodynamical simulations of the Lyman-alpha forest,''
  Mon.\ Not.\ Roy.\ Astron.\ Soc.\  {\bf 365} (2006) 231
  [arXiv:astro-ph/0508177].

\bibitem{viel2006}
  M.~Viel, M.~G.~Haehnelt and A.~Lewis,
  ``The Lyman-alpha forest and WMAP year three,''
  Mon.\ Not.\ Roy.\ Astron.\ Soc.\  {\bf 370} (2006) L51
  [arXiv:astro-ph/0604310].

\bibitem{Lewis:2002ah}
  A.~Lewis and S.~Bridle,
  ``Cosmological parameters from CMB and other data:
  A Monte-Carlo approach,''
  Phys.\ Rev.\ D {\bf 66} (2002) 103511
  [arXiv:astro-ph/0205436]

\bibitem{cosmomc}
  A.~Lewis, Homepage, {\tt http://cosmologist.info}


\bibitem{Zunckel:2006mt}
  C.~Zunckel and P.~G.~Ferreira,
  ``Conservative estimates of the mass of the neutrino from cosmology,''
  arXiv:astro-ph/0610597.

\bibitem{Fogli:2006yq}
  G.~L.~Fogli {\it et al.},
  ``Observables sensitive to absolute neutrino masses: A reappraisal after
  WMAP-3y and first MINOS results,''
  Phys.\ Rev.\  D {\bf 75} (2007) 053001
  [arXiv:hep-ph/0608060].

\bibitem{Cirelli:2006kt}
  M.~Cirelli and A.~Strumia,
  ``Cosmology of neutrinos and extra light particles after WMAP3,''
  JCAP {\bf 0612} (2006) 013
  [arXiv:astro-ph/0607086].

\bibitem{Goobar:2006xz}
  A.~Goobar, S.~Hannestad, E.~M\"ortsell and H.~Tu,
  ``A new bound on the neutrino mass from the SDSS baryon acoustic peak,''
  JCAP {\bf 0606} (2006) 019
  [arXiv:astro-ph/0602155].


\bibitem{Zioutas:2004hi}
  K.~Zioutas {\it et al.}  [CAST Collaboration],
  ``First results from the CERN axion solar telescope (CAST),''
  Phys.\ Rev.\ Lett.\  {\bf 94} (2005) 121301
  [arXiv:hep-ex/0411033].

\bibitem{Andriamonje:2007ew}
  S.~Andriamonje {\it et al.}  [CAST Collaboration],
  ``An improved limit on the axion-photon coupling from the CAST experiment,''
  JCAP {\bf 0704} (2007) 010
  [arXiv:hep-ex/0702006].

\bibitem{CAST2007}
  K.~Zioutas et al.\ [CAST Collaboration],
  ``Status Report of the CAST experiment and request to run
  beyond 2007,''
  CERN-SPSC-2007-013 (5 April 2007), see\\
   http://doc.cern.ch//archive/electronic/cern/preprints/spsc/public/spsc-2007-013.pdf


\bibitem{Hannestad:2005gj}
  S.~Hannestad,
  ``Neutrino masses and the dark energy equation of state:
  Relaxing the cosmological neutrino mass bound,''
  Phys.\ Rev.\ Lett.\  {\bf 95} (2005) 221301
  [arXiv:astro-ph/0505551].

\bibitem{planck}
Planck Bluebook,
{\tt
http://www.rssd.esa.int/index.php?project=Planck}

\bibitem{Cooray:1999rv}
  A.~R.~Cooray,
  ``Weighing neutrinos: Weak lensing approach,''
  Astron.\ Astrophys.\  {\bf 348} (1999) 31
  [arXiv:astro-ph/9904246].

\bibitem{Song:2003gg}
  Y.~S.~Song and L.~Knox,
  ``Dark energy tomography,''
Phys.\ Rev.\ D {\bf 70} (2004)  063510
[arXiv:astro-ph/0312175].

\bibitem{Hannestad:2006as}
  S.~Hannestad, H.~Tu and Y.~Y.~Y.~Wong,
  ``Measuring neutrino masses and dark energy with weak lensing tomography,''
  JCAP {\bf 0606} (2006) 025
  [arXiv:astro-ph/0603019].

\bibitem{Metcalf:2006ji}
  R.~B.~Metcalf and S.~D.~M.~White,
  ``High-resolution imaging of the cosmic mass distribution from gravitational
  lensing of pregalactic HI,''
  arXiv:astro-ph/0611862.

\bibitem{Takada:2005si}
  M.~Takada, E.~Komatsu and T.~Futamase,
  ``Cosmology with high-redshift galaxy survey: Neutrino mass and  inflation,''
  Phys.\ Rev.\  D {\bf 73} (2006) 083520
  [arXiv:astro-ph/0512374].

\bibitem{Hannestad:2007cp}
  S.~Hannestad and Y.~Y.~Y.~Wong,
  ``Neutrino mass from future high redshift galaxy surveys: Sensitivity and
  detection threshold,''
  arXiv:astro-ph/0703031.

\end{thebibliography}
\end{document}